\documentclass[pra,showpacs,preprintnumbers,amsmath,amssymb]{revtex4}
\usepackage{amsmath}

\usepackage{graphicx}
\usepackage{dcolumn}
\usepackage{bm}
\usepackage[                   
            pdfstartview=FitH,
            colorlinks, 
            pdfborder=001,   
            linkcolor=blue,
            anchorcolor=blue,
            citecolor=blue,
            urlcolor=blue
            ]{hyperref}
\usepackage{subfigure}
\usepackage{qcircuit}
\usepackage{booktabs}
\usepackage{CJK}
\usepackage{color}
\usepackage{braket}
\usepackage{listings}

\begin{document}


\title{A scalable tripartite Wigner's friend scenario}

\author{Dong Ding$^{1}$, Can Wang$^{1}$}
\author{Ying-Qiu He$^{1}$}
\email{heyq@ncist.edu.cn}
\author{Tong Hou$^{1}$}
\author{Ting Gao$^{2}$}
\email{gaoting@hebtu.edu.cn}
\author{Feng-Li Yan$^{3}$}
\email{flyan@hebtu.edu.cn}

\affiliation {
$^1$ College of Science, North China Institute of Science and Technology, Beijing 101601, China \\
$^2$ School of Mathematical Sciences, Hebei Normal University, Shijiazhuang 050024, China\\
$^3$ College of Physics, Hebei Normal University, Shijiazhuang 050024, China}
\date{\today}

\begin{abstract}

Wigner's friend thought experiment is intended to reveal the inherent tension between unitary evolution and measurement collapse.
On the basis of Wigner's friend experiment, Brukner derives a no-go theorem for observer-independent facts.
We construct an extended Wigner's friend scenario including three laboratories, namely, Alice's laboratory, Bob's laboratory and Charlie's laboratory, where Alice, Bob and Charlie are standing outside the laboratories while their friends are placed inside their own laboratories. We consider quantum simulation via Q\# quantum programming and also realize the primary quantum circuits using IBM quantum computers. Then, we calculate the probabilities and corresponding statistical uncertainties.
It has been shown that the results of quantum simulation are clearly consistent with theoretical values, while it has a slightly higher
error rates for the experimental results of quantum computers mainly because of a series of quantum gates, especially CNOT gates.

\end{abstract}




\maketitle

\section{Introduction}

In the general logic thinking of human beings, the objective fact does not depend on human will.
Once different observers have different facts, one needs to know whether or not these facts are compatible.
This leads to the debates of the observer-independent facts \cite{Winger1961,Winger-FR2018,Brukner2018,Winger-L2018}.

In the famous ``Schr\"{o}dinger's cat'' thought experiment, the pitiful cat in the box faces life or death contingent upon an automatic device related to an atomic state. In classical physics, when the box is opened the cat must be alive or dead, alternatively. Quantum mechanically, however, as long as one does not open the box the cat's life and death is in an uncertain superposition state.
In 1961, Eugene Wigner \cite{Winger1961} proposed another argument concerning the experimental test of quantum theory beyond  microscopic domains, known as the ``Wigner's friend'' thought experiment. Wigner's friend experiment is the variation on Schr\"{o}dinger's cat and a comprehensible statement is as follows.
Suppose there is a quantum state $(|0\rangle+|1\rangle)/\sqrt{2}$. Wigner's friend, in an isolated laboratory, measures this state in \{$|0\rangle,|1\rangle$\} basis and then the outcome is either $|0\rangle$ or $|1\rangle$.
Wigner is standing outside the laboratory and only knows that his friend has measured the state, but he does not know the specific measurement result.
In view of the unitary evolution of closed systems \cite{NC2000,Quantum-entanglement,Entanglement-detection2009}, Wigner therefore describes the composite system including the state and friend's record as an entangled state, i.e.,
$(|0\rangle|{\rm ``record~is~0"}\rangle +
  |1\rangle|{\rm ``record~is~1"}\rangle)/\sqrt{2}$.
Once Wigner has verified that the record of friend and the quantum state are indeed superimposed through interference experiments, he can conclude that friend must not record a clear result. However, friend can always declare that he/she had recorded a certain result.
Then, how can one reconcile the different accounts of these two processes.

More recently, Caslav Brukner \cite{Brukner2018} proposed a no-go theorem for observer-independent facts by constructing an extended Wigner's friend scenario with four observers. The no-go theorem shows that the following four statements ``universal validity of quantum theory'', ``locality'', ``freedom of choice'', and ``observer-independent facts'' are incompatible, and thus if one holds the first three statements then it is impossible to reconcile the observed outcomes of different observers.
In a state-of-the-art six-photon experiment, Proietti \emph{et al} \cite{Proietti2019} realized this extended Wigner's friend scenario by using three pairs of entangled photons and two optical fusion gates \cite{Fusion2005}.
It convincingly demonstrates the incompatibility of these statements in Brukner's no-go theorem with the aid of the violation of the Clauser-Horne-Shimony-Holt inequality \cite{CHSH1969}.
For the Wigner's friend experiment involving multipartite laboratories, in view of the Greenberger-Horne-Zeilinger (GHZ) theorem Brukner \cite{Brukner2018} and Leegwater \cite{Winger-L2018} respectively proposed tripartite Wigner's friend experiments in a way without inequalities; however, to our knowledge the Wigner's friend scenario including the multipartite inequality has not yet been shown.

In this paper, we propose a tripartite Wigner's friend scenario with the central source W state and three pair ancillary Bell states.
Using a generalized tripartite correlation inequality, after three fusion gates the remaining six particles are capable of verifying the Brukner's no-go theorem in the case of three laboratories.
Then, we use Q\# quantum program to simulate the present scenario and the results show that the inequality is clearly violated.
This means that quantum programming is effective as a tool for studying the behaviors of the multipartite quantum systems.
Also, we provide experimental realizations of two primary quantum circuits using IBM quantum computers.

\section{The tripartite Wigner's friend scenario}

Consider three separate laboratories controlled by Alice, Bob and Charlie, respectively. Alice, Bob and Charlie are outside the laboratories, and meanwhile, each of them has a friend inside their own laboratory.
We assume that Alice can freely choose measurement settings described by variables $A_{0}$ or $A_{1}$, and similarly, $B_{0}$ or $B_{1}$, and $C_{0}$ or $C_{1}$ for Bob and Charlie respectively.
Wherein, $A_{0}$, $B_{0}$ and $C_{0}$ correspond to the records of measurement results for Alice's friend, Bob's friend and Charlie's friend, while $A_{1}$, $B_{1}$ and $C_{1}$ are respectively associated with joint measurements of the composite system including the state and friend's record.
For simplicity, take values $v_1$, $v_2$ and $v_3$ to be $+1$ or $-1$ for $A_{k_{1}}$, $B_{k_{2}}$ and $C_{k_{3}}$, $k_{1}, k_{2}, k_{3}=0,1$, respectively.
Moreover, assume that the joint probability $P(A_0,A_1,B_0,B_1,C_0, C_1)$ whose marginals match the probabilities $P(A_{k_{1}},B_{k_{2}},C_{k_{3}})$ satisfies the generalized tripartite correlation inequality \cite{DHYG2020}
\begin{eqnarray}\label{inequality_I}
\mathcal{I} = \frac{1}{8}\sum \limits_{j_{1},j_{2},j_{3}=0,1}| \sum\limits_{k_{1},k_{2},k_{3}=0,1}(-1)^{\vec{k}\cdot\vec{j}}E_{k_{1}k_{2}k_{3}} |\leq 1,
\end{eqnarray}
where $E_{k_{1}k_{2}k_{3}}=\langle A_{k_{1}}B_{k_{2}}C_{k_{3}}\rangle =\sum_{v_1,v_2,v_3}v_1v_2v_3 P(A_{k_{1}}=v_1,B_{k_{2}}=v_2,C_{k_{3}}=v_3)$.

Suppose that three laboratories initially share an entangled state
\begin{eqnarray}\label{input-state}
|\psi\rangle_{abc}=\frac{1}{\sqrt{3}}[\cos\theta(|001\rangle+|010\rangle-|100\rangle)+\sin\theta(|000\rangle+|101\rangle+|110\rangle)],
\end{eqnarray}
and particles $a$, $b$ and $c$ belong to Alice's, Bob's and Charlie's laboratories, respectively.
This state can be directly prepared from W state, i.e., $|W\rangle=(|001\rangle+|010\rangle+|100\rangle)/\sqrt{3}$, by applying unitary operator $U_{\theta}=\cos\theta \sigma_{z}+\sin\theta\sigma_{x}$ on particle $a$, where $\sigma_{x}$ and $\sigma_{z}$ are the Pauli operators.

\begin{figure}
  \centering\includegraphics[width=4.0in]{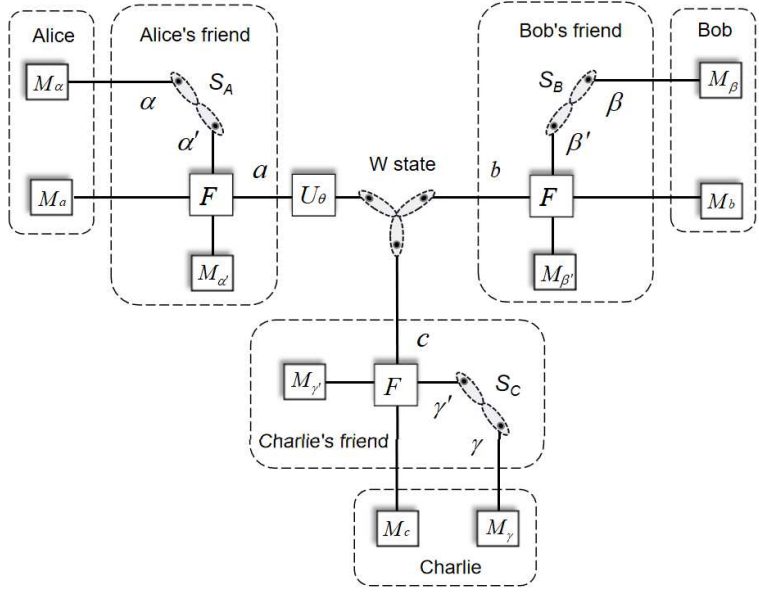}\\
  \caption{The schematic diagram of the tripartite Wigner's friend scenario.}
  \label{schematic}
\end{figure}

Just as bipartite Wigner's friend experiment \cite{Proietti2019}, to characterize the individual facts for different observers we here introduce a pair of auxiliary entangled particles in each laboratory.
Then, choose three pairs of Bell states $|\psi^{-}\rangle_{XX'}=(|01\rangle_{XX'}-|10\rangle_{XX'})/\sqrt{2}$ with $X=\alpha,\beta,\gamma$ for Alice's, Bob's and Charlie's laboratories, respectively, as shown in Fig.\ref{schematic}.
So the combined system including nine particles is initially described by the state
\begin{eqnarray}
|\psi_{\rm in}\rangle=|\psi\rangle_{abc}|\psi^{-}\rangle_{\alpha \alpha'}|\psi^{-}\rangle_{\beta \beta'}|\psi^{-}\rangle_{\gamma \gamma'}.
\end{eqnarray}
For simplicity, we henceforth call particles $a,b,c$ as the signal particles and $\alpha,\alpha',\beta,\beta',\gamma,\gamma'$ as the auxiliary particles.

In each laboratory, one would extract a piece of information of the signal particle and then store it in an auxiliary particle, corresponding to the friend's record (or fact).
This procedure can be described by fusion gate \cite{Proietti2019,Fusion2005}, described by
\begin{eqnarray}
  F = \frac{1}{\sqrt{2}}(|0\rangle\langle0|\langle0|-|1\rangle\langle1|\langle1|).
\end{eqnarray}
More specifically, we take Alice's laboratory for example and consider fusion gate $(|0\rangle_a\langle0|_a\langle0|_{\alpha'}-|1\rangle_a\langle1|_a\langle1|_{\alpha'})/\sqrt{2}$. Alice's friend, who initially holds particles $a,\alpha,\alpha'$ inside the laboratory, is about to perform a measurement on the auxiliary particle $\alpha'$ to determine whether particles $a$ and $\alpha'$ is in $|0\rangle_a|0\rangle_{\alpha'}$ or $|1\rangle_a|1\rangle_{\alpha'}$.
Similarly, the fusion gates $(|0\rangle_b\langle0|_b\langle0|_{\beta'}-|1\rangle_b\langle1|_b\langle1|_{\beta'})/\sqrt{2}$
and $(|0\rangle_c\langle0|_c\langle0|_{\gamma'}-|1\rangle_c\langle1|_c\langle1|_{\gamma'})/\sqrt{2}$
correspond to Bob's and Charlie's laboratories, respectively.
The essence of the fusion gates is to provide observers Alice, Bob and Charlie with measurement settings $A_0,A_1,B_0,B_1,C_0,C_1$ which can be used for test of the observer-independent facts through the violation of appropriate correlation inequalities.

After carrying out the fusion gates, the renormalized six-particle state becomes
\begin{eqnarray}\label{six-particle}
|\psi\rangle_{a\alpha b\beta c\gamma}=\frac{1}{\sqrt{3}}[\cos\theta(|010110\rangle+|011001\rangle-|100101\rangle) +\sin\theta(|100110\rangle+|101001\rangle+|010101\rangle)].
\end{eqnarray}
To proceed, outside the laboratories Alice, Bob and Charlie are free to perform either \{$|0\rangle,|1\rangle$\} basis or Bell basis measurement on their own two particles. Each \{$|0\rangle,|1\rangle$\} basis measurement, $A_0,B_0$ or $C_0$ reveals his/her friend's record; while the Bell basis measurement, $A_1,B_1$ or $C_1$ corresponds to the fact from Alice, Bob or Charlie, respectively.
Obviously there are eight possible measurement settings, and a straightforward but lengthy calculation shows that
\begin{eqnarray}\label{E12}
E_{000} = -\cos2\theta, ~~ E_{100} = -\sin2\theta,
\end{eqnarray}
\begin{eqnarray}\label{E345}
E_{010} = E_{001} = E_{111} = \frac{2}{3}\sin2\theta,
\end{eqnarray}
\begin{eqnarray}\label{E678}
E_{011} = -E_{110} = -E_{101} = \frac{2}{3}\cos2\theta.
\end{eqnarray}
Substituting these results into the left hand side of the correlation inequality (\ref{inequality_I}) yields
\begin{eqnarray}{}
\mathcal{I} &=& \frac{1}{8}[|\sin2\theta-\frac{5}{3}\cos2\theta| + |\sin2\theta+\frac{5}{3}\cos2\theta|
             +|\cos2\theta+\frac{5}{3}\sin2\theta| + |\cos2\theta-\frac{5}{3}\sin2\theta|  \nonumber \\
        & &  +\frac{10}{3}|\cos2\theta+\sin2\theta|+ \frac{10}{3}|\cos2\theta-\sin2\theta|].
\end{eqnarray}
A simple calculation shows that the maximal quantum prediction of this correlation polynomial is $\mathcal{I} = 1.5$ with $\theta=0$ or integer multiples of
$\pi/4$.

Wigner's friend thought experiment focuses on two postulates of quantum mechanics, unitary evolution and wavefunction collapse.
That is, for an observer outside the laboratory, any closed quantum system may be described by a unitary evolution; while for the observer inside the laboratory, if the quantum system is measured it will collapse into an eigenstate of the measurement operator.
In light of the Brukner's no-go theorem \cite{Brukner2018}, the different facts of the observers inside and outside the laboratories can be manipulated by converting to the corresponding measurement settings related to correlation inequalities. Then any violation of the inequalities may indicate the contradiction between the assumptions in no-go theorem.
Hence, the present scenario may provide a useful means of testing the local observer independence, and also it provides evidence for
verifying the no-go theorem in the case of three laboratories.

\section{Quantum simulation and experiments}

\subsection{Quantum circuit}

We now construct a quantum circuit to realize the present scenario. The quantum circuit mainly includes preparation of the entangled states, fusion gates, and quantum measurements outside the laboratories.

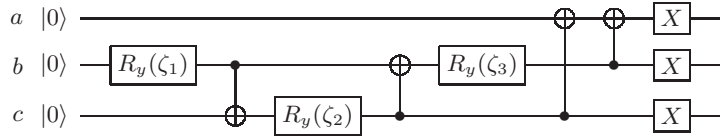
\begin{figure}[htb]
\centerline{
\Qcircuit @C=1.2em @R=0.5em {
& \lstick{a~~} & \lstick{\ket{0}} & \qw      &\qw      &\qw      &\qw      &\qw      &\targ    &\targ    &\gate{X}  &\qw  &\\
& \lstick{b~~} & \lstick{\ket{0}} & \gate{R_{y}(\zeta_{1})} &\ctrl{1} &\qw      &\targ    &\gate{R_{y}(\zeta_{3})} &\qw      &\ctrl{-1}&\gate{X} &\qw    &\\
& \lstick{c~~} & \lstick{\ket{0}} & \qw      &\targ    &\gate{R_{y}(\zeta_{2})} &\ctrl{-1}&\qw      &\ctrl{-2}&\qw      &\gate{X} &\qw  &\\
}}
\vskip 0.55\baselineskip
\centerline{\footnotesize}
\caption{The quantum circuit diagram of preparation of the W state using rotation operators $R_{y}(\zeta)$.}
\label{W1}
\vskip 0.55\baselineskip
\end{figure}

\begin{figure}[htb]
\centerline{
\Qcircuit @C=1.2em @R=0.5em {
& \lstick{a~~} & \lstick{\ket{1}} &\qw                &\qw      &\ctrl{1} &\qw      &\qw                &\targ     &\qw      &\qw
&\qw                &\qw        &\qw  &\\
& \lstick{b~~} & \lstick{\ket{0}} &\gate{U({\xi_{1}})} &\gate{H} &\targ    &\gate{H} &\gate{U({\xi_{1}})} &\ctrl{-1} &\ctrl{1} &\qw
&\qw                &\targ      &\qw  &\\
& \lstick{c~~} & \lstick{\ket{0}} &\qw                &\qw      &\qw      &\qw      &\gate{U({\xi_{2}})} &\gate{H}  &\targ    &\gate{H}  &\gate{U({\xi_{2}})} &\ctrl{-1}  &\qw  &\\
}}
\vskip 0.55\baselineskip
\centerline{\footnotesize}
\caption{The quantum circuit diagram of preparation of the W state using unitary operators $U({\xi})$.}
\label{W2}
\vskip 0.55\baselineskip
\end{figure}
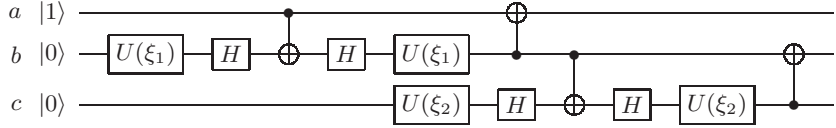

Firstly, consider the task of preparing the W state.
We here provide two methods to achieve this \cite{BBHB1997,LongS2001,ZWL-W-state,WDiker2016}. One is using rotation operators $R_{y}(\zeta)$ and the other is to use unitary operators $U({\xi})$, each of them needs four CNOT gates.
Consider the rotation operator about the $y$ axis with the equation
\begin{eqnarray}
R_{y}(\zeta) = {\rm e}^{{\rm -i}\zeta \sigma_{y}/2} =
\left(
  \begin{array}{cc}
    \cos \frac{\zeta}{2} & -\sin \frac{\zeta}{2}\\
    \sin \frac{\zeta}{2} & \cos \frac{\zeta}{2}\\
  \end{array}
\right).
\end{eqnarray}
As shown in Fig.\ref{W1}, consider three signal particles $a,b,c$ in the initial state $|000\rangle_{abc}$ and take $\zeta_{1} = \zeta_{3} = 2\arcsin\sqrt{(5+\sqrt{5})/10}$, $\zeta_{2} = -2\arcsin\sqrt{(3-\sqrt{5})/6}$. Then, after an $X$ gate (NOT gate) on each signal particle, the signal particles can be prepared in W state.
On the other hand, considering an input state in the initial state $|100\rangle_{abc}$ and using the unitary operator
\begin{eqnarray}
U(\xi) =
\left(
  \begin{array}{cc}
    \cos \frac{\xi}{2} &  \sin \frac{\xi}{2}\\
    \sin \frac{\xi}{2} & -\cos \frac{\xi}{2}\\
  \end{array}
\right),
\end{eqnarray}
as shown in Fig.\ref{W2}, take $\xi_{1} = \arccos (1/\sqrt{3})$ and $\xi_{2} = \pi/4$, one can also prepare the signal particles in W state.

To proceed, applying a unitary operator $U_{\theta}$ to the signal particle $a$, then one can obtain the state (\ref{input-state}).
Also, we set the remaining auxiliary particles $\alpha,\alpha',\beta,\beta',\gamma,\gamma'$ to {\it one}, and three pairs of Bell states can be easily obtained by using $H$ gates and CNOT gates. By now, we complete the preparation of the initial state.

A crucial element for quantum circuit is the fusion gate, which includes two particles $a$ and $\alpha'$, or $b$ and $\beta'$, or $c$ and $\gamma'$.
The function of the fusion gate is to determine whether these two particles are in the state $|00\rangle$ or $|11\rangle$, but not distinguishing between them.
To do this, we consider a CNOT gate followed by a measurement on the auxiliary particle $\alpha'$, $\beta'$ or $\gamma'$. Take Alice's laboratory for example. The CNOT gate can evolve the particles $a$ and $\alpha'$ by the following transformations:
\begin{eqnarray}
|0\rangle_a |0\rangle_{\alpha'} \rightarrow |0\rangle_a |0\rangle_{\alpha'}, ~~
|0\rangle_a |1\rangle_{\alpha'} \rightarrow |0\rangle_a |1\rangle_{\alpha'}, ~~
|1\rangle_a |0\rangle_{\alpha'} \rightarrow |1\rangle_a |1\rangle_{\alpha'}, ~~
|1\rangle_a |1\rangle_{\alpha'} \rightarrow |1\rangle_a |0\rangle_{\alpha'}.
\end{eqnarray}
In this way, the states $|0\rangle_a |0\rangle_{\alpha'}$ or $|1\rangle_a |1\rangle_{\alpha'}$ can be postselected by choosing the measurement result $|0\rangle_{\alpha'}$.
Similar to the cases for Bob's and Charlie's laboratories, thus again, one can obtain the six-particle state (\ref{six-particle}) with probability $1/8$, condition on the measurement result of the auxiliary particles is $|000\rangle_{\alpha'\beta'\gamma'}$.

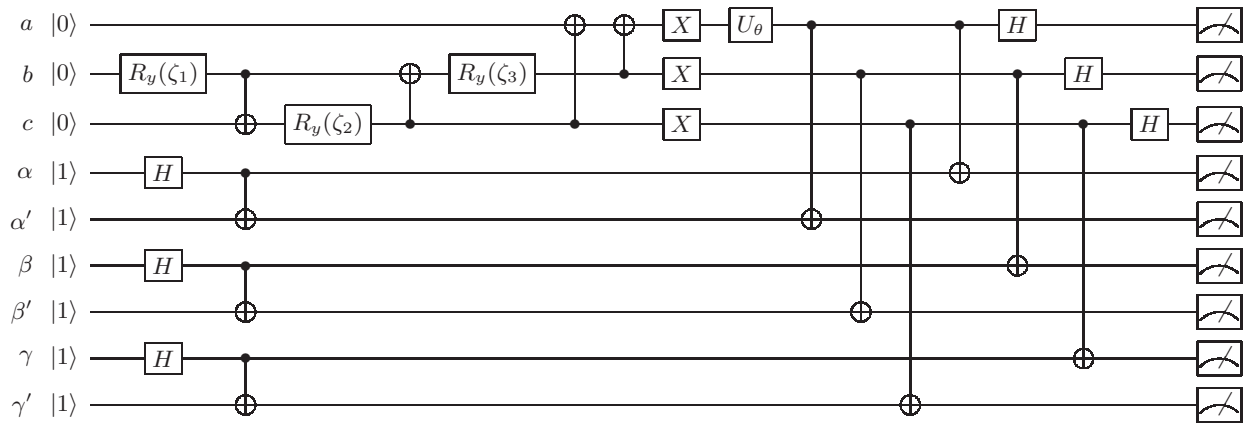
\begin{figure}[htb]
\centerline{
\Qcircuit @C=1.2em @R=0.5em {
& \lstick{a~~} & \lstick{\ket{0}} & \qw      &\qw      &\qw      &\qw      &\qw      &\targ    &\targ    &\gate{X} &\gate{U_{\theta}}          &\ctrl{4}    &\qw      &\qw   &\ctrl{3}  & \gate{H}  &\qw       &\qw     &\meter & \\
& \lstick{b~~} & \lstick{\ket{0}} & \gate{R_{y}(\zeta_{1})} &\ctrl{1} &\qw      &\targ    &\gate{R_{y}(\zeta_{3})} &\qw      &\ctrl{-1}&\gate{X} &\qw     &\qw     &\ctrl{5 }    &\qw      &\qw    &\ctrl{4}  & \gate{H}  &\qw      &\meter  & \\
& \lstick{c~~} & \lstick{\ket{0}} & \qw      &\targ    &\gate{R_{y}(\zeta_{2})} &\ctrl{-1}&\qw      &\ctrl{-2}&\qw      &\gate{X} &\qw               &\qw      &\qw      &\ctrl{6}     &\qw   &\qw    &\ctrl{5}  & \gate{H}   &\meter  &  \\
& \lstick{\alpha~~} & \lstick{\ket{1}} & \gate{H} &\ctrl{1} &\qw      &\qw      &\qw      &\qw      &\qw      &\qw      &\qw               &\qw      &\qw      &\qw      &\targ       &\qw    &\qw   &\qw       &\meter  & \\
& \lstick{\alpha{'}~~} & \lstick{\ket{1}} & \qw      &\targ    &\qw      &\qw      &\qw      &\qw      &\qw      &\qw      &\qw          &\targ    &\qw      &\qw        &\qw     &\qw     &\qw     &\qw      &\meter&  \\
& \lstick{\beta~~}  & \lstick{\ket{1}} & \gate{H} &\ctrl{1} &\qw      &\qw      &\qw      &\qw      &\qw      &\qw      &\qw               &\qw      &\qw      &\qw       &\qw     &\targ     &\qw    &\qw     &\meter & \\
& \lstick{\beta{'}~~} & \lstick{\ket{1}} & \qw      &\targ    &\qw      &\qw      &\qw      &\qw      &\qw      &\qw      &\qw           &\qw      &\targ    &\qw         &\qw    &\qw     &\qw     &\qw       &\meter & \\
& \lstick{\gamma~~} & \lstick{\ket{1}} & \gate{H} &\ctrl{1} &\qw      &\qw      &\qw      &\qw      &\qw      &\qw      &\qw              &\qw      &\qw      &\qw         &\qw   &\qw    &\targ     &\qw        &\meter & \\
& \lstick{\gamma{'}~~} & \lstick{\ket{1}} & \qw      &\targ    &\qw      &\qw      &\qw      &\qw      &\qw      &\qw      &\qw         &\qw      &\qw      &\targ        &\qw   &\qw     &\qw     &\qw       &\meter& \\
}}
\vskip 0.55\baselineskip
\centerline{\footnotesize}
\caption{A quantum circuit diagram of the settings $A_1B_1C_1$ with rotation operators for the tripartite Wigner's friend scenario.}
\label{circuit}
\vskip 0.55\baselineskip
\end{figure}

In each laboratory, Alice can choose to measure either $A_0$ or $A_1$, Bob chooses to measure either $B_0$ or $B_1$, and Charlie chooses to measure either $C_0$ or $C_1$. In quantum circuit, $A_0$, $B_0$ and $C_0$ indicate the \{$|0\rangle,|1\rangle$\} basis measurements; $A_1$, $B_1$ and $C_1$ are Bell basis measurements, and each of them can be described using a CNOT gate and a H gate followed by the \{$|0\rangle,|1\rangle$\} basis measurement.
By this, we can construct the quantum circuits for the present scenario.
As an example, consider a quantum circuit of the settings $A_1B_1C_1$ shown in Fig.\ref{circuit}.

\subsection{Quantum simulation via quantum programming}

We next provide quantum simulation of the present scenario via quantum programming \cite{Q-sharp2018,Quantum-programming-languages2020,HDWZH2021}.
Consider Q\# quantum programming language in {\it Visual Studio}, an integrated development environment, developed by Microsoft.

Q\# is a multiparadigm quantum programming language that nearly all operations required for quantum computing can be well defined.
In {\it Visual Studio}, one can create a Q\# Application which contains two files: one is {\it Operations.qs}, compiling and debugging quantum program in Q\#, and the other is {\it Driver.cs}, writing classically-controlled program in C\#.
To demonstrate the maximum quantum violation, we here take $\theta=\pi/4$ and then write the programs on the basis of the quantum circuit.
There are eight groups of measurement settings, i.e.,
$A_{0}B_{0}C_{0}$, $A_{1}B_{0}C_{0}$, $A_{0}B_{1}C_{0}$, $A_{0}B_{0}C_{1}$, $A_{1}B_{1}C_{0}$, $A_{1}B_{0}C_{1}$, $A_{0}B_{1}C_{1}$, $A_{1}B_{1}C_{1}$,
and consequently we consider eight {\it operations} and corresponding classical controls.
For each {\it operation} we set a loop of 10,000 steps, and all of the {\it operations} will be executed 5 times.
The average values of measurement results and the corresponding expectation values of the observables for these eight groups of measurement settings are shown in Table \ref{Table1}.

\begin{table}[!htbp]
\caption{The average values of measurement results and the expectation values of the observables for eight groups of measurement settings.}
\label{Table1}
{\begin{tabular}{c p{1cm}<{\centering}p{1cm}<{\centering}p{1cm}<{\centering}p{1cm}<{\centering}p{1cm}<{\centering}p{1cm}<{\centering}p{1cm}<{\centering}p{1cm}<{\centering}p{2.8cm}<{\centering}}
\specialrule{0.1em}{3pt}{1pt}
\quad & +~+~+~~~ & +~+~--~~~ & +~--~+~~~ & +~--~--~~~ & --~+~+~~~ & --~+~--~~~ & --~--~+~~~ & --~--~--~~~ & Expectation value \\
\specialrule{0.1em}{1pt}{1pt}
$ A_0B_0C_0 $ & 1696 &$ 1642 $&$ 1675 $&$ 0    $&$ 1627 $&$ 1695 $&$ 1659 $&$ 0    $&$ 0.0106   $ \\
$ A_1B_0C_0 $ & 0    &$ 3323 $&$ 3314 $&$ 0    $&$ 3364 $&$ 0    $&$ 0    $&$ 0    $&$ -1.0001  $ \\
$ A_0B_1C_0 $ & 3329 &$ 837  $&$ 0    $&$ 818  $&$ 0    $&$ 845  $&$ 3321 $&$ 851  $&$ 0.6625   $ \\
$ A_0B_0C_1 $ & 3327 &$ 0    $&$ 845  $&$ 837  $&$ 0    $&$ 3345 $&$ 808  $&$ 838  $&$ 0.6634   $ \\
$ A_1B_1C_0 $ & 1673 &$ 1687 $&$ 1659 $&$ 1653 $&$ 1658 $&$ 0    $&$ 1670 $&$ 0    $&$ -0.0008  $ \\
$ A_1B_0C_1 $ & 1650 &$ 1680 $&$ 1685 $&$ 1696 $&$ 1654 $&$ 1635 $&$ 0    $&$ 0    $&$ -0.0038  $ \\
$ A_0B_1C_1 $ & 3746 &$ 427  $&$ 410  $&$ 425  $&$ 407  $&$ 426  $&$ 425  $&$ 3735 $&$ 0.0043   $ \\
$ A_1B_1C_1 $ & 3347 &$ 0    $&$ 0    $&$ 3328 $&$ 826  $&$ 845  $&$ 829  $&$ 824  $&$ 0.6699   $ \\
\specialrule{0.1em}{1pt}{3pt}
\end{tabular}}
\end{table}

\begin{figure}
  \centering\includegraphics[width=7.0in]{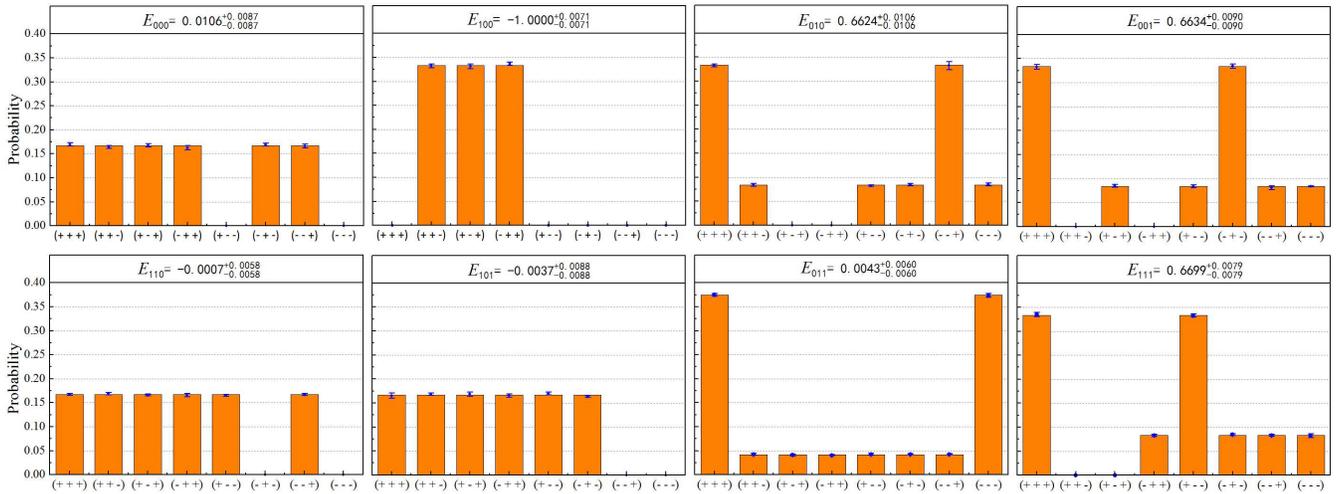}\\
  \caption{(color online). The outcome probabilities of the expectation values for the eight groups of observables.}
  \label{E8}
\end{figure}

According to the statistical measurement results, the count difference between different running is actually very small.
We draw the outcome probabilities comprising each of the eight expectation values, as shown in Fig.\ref{E8}.
The signs ``$+$'' and ``$-$'' on the horizontal axis mean that the eigenvalues of the corresponding observables are $+1$ and $-1$, respectively. Orange bars are the theoretical predictions and error bars indicate statistical errors.
Finally, using error propagation we calculate the statistical uncertainties of the correlation polynomial, and then we have
\begin{eqnarray}
\mathcal{I} = 1.498 \pm 0.009.
\end{eqnarray}
Obviously, the result is very close to the theoretical value $1.5$. This means that the present programmable quantum circuit is feasible to construct the scalable tripartite Wigner's friend experiment in quantum computers.

\subsection{Experiments}

\begin{figure}
  \centering\includegraphics[width=2.6in]{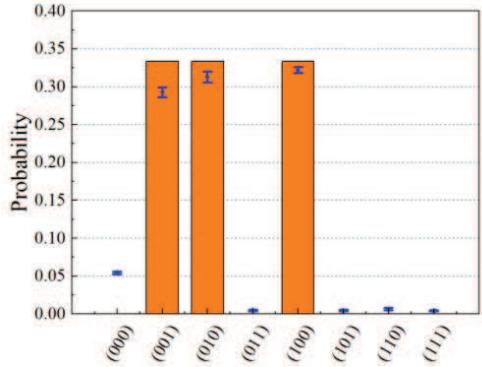}\\
  \caption{(color online). The outcome probabilities for preparing the W state using the {\it ibmq\_lima} device.}
  \label{PW1}
\end{figure}

\begin{figure}
  \centering\includegraphics[width=6.0in]{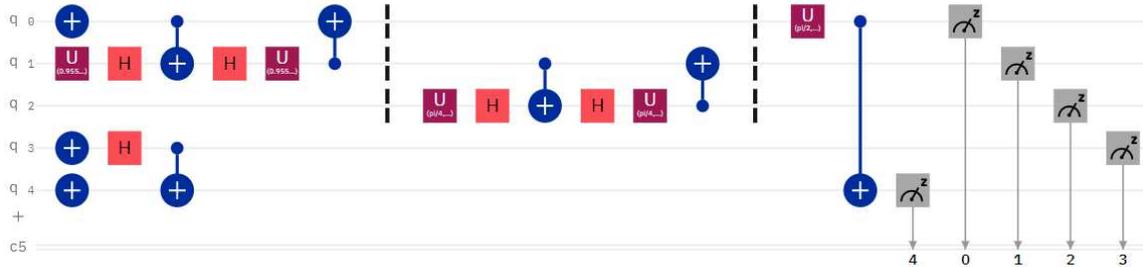}\\
  \caption{(color online). The quantum circuit realizing the fusion gate for Alice's laboratory.}
  \label{Fusion}
\end{figure}

IBM offers a cloud platform to carry out quantum calculation relying on their superconducting quantum computers, where the quantum circuits can be built through the IBM Quantum Lab with Qiskit \cite{IBM,SDJ2021,BGT2021IBM}. There are 20 quantum systems consisting of up to 65 qubits, and currently six 5-qubit devices of them are publicly available.
By this, we here experimentally realize two primary quantum circuits, preparing the W state and demonstrating the fusion gate, using IBM quantum computers.
We run all of these 5-qubit devices and then consider the {\it ibmq\_lima} and {\it ibmq\_belem} devices for our choice because of their error rates are relatively smaller than many of the others. We repeat the experiments 5 times and each circuits are implemented in 8192 shots.
For preparation of the W state, we test the quantum circuit shown in Fig.\ref{W2} using the {\it ibmq\_lima} device, and then calculate the probabilities of all basis states and corresponding statistical uncertainties. The results of the experiments are basically consistent with theoretical values, as shown in Fig. \ref{PW1}.
For demonstration of the fusion gate, due to the limitation of the available qubits we here take Alice's laboratory for example. Consider a composite system consisting of three signal particles initially in W state and the auxiliary Bell state in Alice's laboratory.
The quantum circuit of this fusion gate is composed of a CNOT gate, setting particle $a$ as control qubit and particle $\alpha'$ as target qubit, and the followed measurement on the particle $\alpha'$.
Condition on collecting the result $|0\rangle_{\alpha'}$, the fusion gate is capable of providing $A_0$ as the fact of Alice's friend and also providing $A_1$ as Alice's fact, both of which are dependent on the final choice of measurement settings by Alice outside the laboratory.
We consider the {\it ibmq\_belem} device and the corresponding quantum circuit are shown in Fig.\ref{Fusion}.

\begin{figure}
  \centering\includegraphics[width=3.5in]{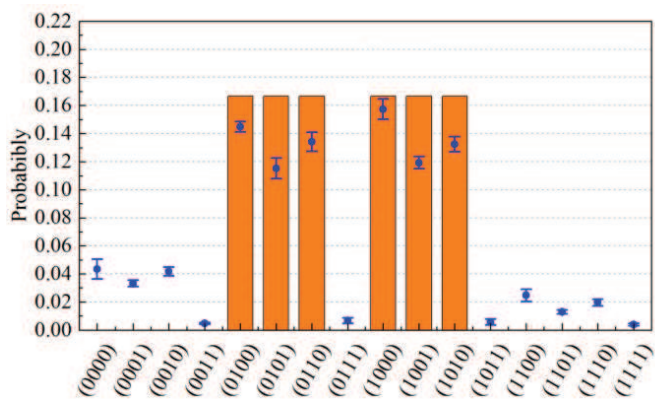}\\
  \caption{(color online). The outcome probabilities for demonstrating the fusion gate in Alice's laboratory using the  {\it ibmq\_lima} device.}
  \label{PF1}
\end{figure}

Theoretically, after performing the fusion gate we renormalize the remaining state condition on the auxiliary particle $\alpha'$ being zero, and then calculate the probabilities of the respective measurement outcomes.
Experimentally, by taking the valid shots (collected roughly in 4130) into account, we calculate the probabilities of measurement results and the corresponding statistical uncertainties, as shown in Fig.\ref{PF1}.
It has a slightly higher error rate for the experimental results of the fusion gate mainly because of a series of quantum gates in quantum circuit, especially CNOT gates. After all, any CNOT gate in quantum device will accumulate a two-qubit gate errors related to connectivity map, respectively.

\section{Discussion and summary}

In summary, we have proposed a tripartite Wigner's friend scenario and its demonstration. By replacing the central Bell state with W state, we extend the Wigner's friend scenario from two laboratories to three laboratories.
Since we consider a multipartite inequality in our scenario, it is different from the previous ``three Wigners and three friends'' schemes \cite{Brukner2018,Winger-L2018}, which refer to the ``without inequality'' GHZ theorem. The violation of the inequality means that the Brukner's no-go theorem holds in the case of three laboratories.
Accordingly, and just as \cite{Proietti2019}, if one adheres to the assumptions of universal validity of quantum theory, locality and freedom of choice, then the recorded facts for different observers may not be reconciled directly.

To illustrate the present scenario, we have designed a programmable quantum circuit consisting of preparation of the initial states, unitary evolution and quantum measurement. The key part of the quantum circuit is realization of the fusion gate. To do this, we introduce an ancilla CNOT gate and followed by \{$|0\rangle,|1\rangle$\} basis measurement, by which it is capable of constructing those measurement settings related to different facts.
According to the quantum circuit we wrote quantum programs in Q\# to simulate our scenario, and furthermore, reported the experimental results for the primary quantum circuits using IBM quantum computers.
Harvesting these statistical values, we conclude that our quantum circuit is feasible to realize the scenario, especially the circuit of realization of the fusion gate is simple and novel.

It is interesting to look for the models for extending the Wigner's friend experiments based on multipartite correlation inequalities \cite{Bell1964,WW2001,ZB2002,Bell-nonlocality2014}.
Note that not all multipartite inequalities are suitable to construct the Wigner's friend experiment.
For example, by extending the Wigner's friend scenario with GHZ state involving two arbitrary single-qubit unitary transformations, the well known Werner-Wolf-Zukowski-Brukner inequality \cite{WW2001,ZB2002} would not be violated.
Although many fundamental challenges remain for real experiment as multiphoton interference in quantum optical systems and manipulation with high precision for superconducting qubits, in any event, investigating the Wigner's friend experiments may promise to stimulate exciting and unexpected links among those postulates of quantum mechanics such as unitary evolution, measurement collapse, and so on.
Finally, we hope that this work will motivate further developments on the fascinating topic of the multipartite Wigner's friend scenarios.

\begin{acknowledgements}
This work was supported by the National Natural Science Foundation of China under Grant Nos: 12071110, 11547169,
the Education Department of Hebei Province Natural Science Foundation under Grant Nos: ZD2021407, ZD2021066, ZD2020167, QN2019305,
the Hebei Natural Science Foundation of China under Grant No: A2020205014,
the Fundamental Research Funds for the Central Universities of Ministry of Education of China under Grant Nos: 3142019020, 3142017069, 050201030507.
\end{acknowledgements}

\end{document}